# *Observation of Cooperative Electronic Quantum Tunneling: Increasing Accessible Nuclear States in a Molecular Qudit*

*Eufemio Moreno-Pineda\*,a, Svetlana Klyatskaya[a], Ping Du[a], Marko Damjanović[a], Gheorghe Taran[b], Wolfgang Wernsdorfer\*,a,b,c and Mario Ruben\*,a,d*

  a.  Institute of Nanotechnology (INT), Karlsruhe Institute of Technology (KIT), Hermann-von-Helmholtz-Platz 1, D-76344 Eggenstein-Leopoldshafen, Germany.
  b.  Physikalisches Institut, Karlsruhe Institute of Technology, D-76131 Karlsruhe, Germany.
  c.  Institut Néel, CNRS, Université Grenoble Alpes, 25 rue des Martyrs, F-38000 Grenoble, France.
  d.  Institut de Physique et Chimie des Matériaux de Strasbourg (IPCMS), CNRS, Université de Strasbourg, 23 rue du Loess, BP 43, F-67034 Strasbourg Cedex 2, France.

**ABSTRACT:** As an extension of two-level quantum bits (qubits), multilevel systems, so-called qu*d*its, where *d* represents the Hilbert space dimension, have been predicted to reduce the number of iterations in quantum computation algorithms. This has been tested in the well-known $[TbPc_2]^0$ SMM, which allowed the implementation of the Grover algorithm in a single molecular unit. In the quest for molecular systems possessing an increased number of accessible nuclear spin states, we explore herein a dimeric $Tb_2$-SMM *via* single crystal μ-SQUID measurements at sub-Kelvin temperatures. We observe ferromagnetic interactions between the Tb(III) ions and cooperative quantum tunneling of the electronic spins with spin ground state $|J_z = \pm 6\rangle$. The strong hyperfine coupling with the Tb(III) nuclear spins leads to a multitude of spin reversal paths leading to seven strong hyperfine driven tunneling steps in the hysteresis loops. Our results show the possibility to read-out the Tb(III) nuclear spin states *via* the cooperative tunneling of the electronic spins, making the dimeric $Tb_2$-SMM an excellent nuclear spin qu*d*it candidate with *d* =16.

*On occasion of the 60[th] anniversary of Prof. Kim Dunbar.*

## INTRODUCTION

The exploitation of the quantum properties in device applications has boosted numerous studies of molecules exhibiting slow relaxation of the magnetization, termed Single-Molecule Magnets (SMMs).[1-6] Moreover, these systems have shown a variety of quantum effects such as the quantum tunneling of the magnetization (QTM)[3], Berry phase[4], quantum oscillations[5] and entanglement.[6] In addition, SMMs systems have also been proposed as parts of quantum computers (QCs) acting as quantum bits (so-called qubits).[7]

Amongst the numerous SMMs, the family of terbium(III) bis(phthalocyaninato) ([TbPc$_2$]$^{0,\pm1}$) complexes have been proposed and established as qubits.[8,9] The electronic properties of [TbPc$_2$]$^0$ permitted the integration of a single molecule in a transistor circuit allowing the initialization, manipulation, and read-out of the nuclear spin states.[9] In particular, in the [TbPc$_2$]$^0$ complex, the Ising-like magnetic anisotropy isolates the ground doublet state $|J_z = \pm 6\rangle$ from excited states, while the hyperfine (*hf*) interaction couples the electronic spin to the nuclear spin ($|I = 3/2\rangle$) of the $^{159}$Tb(III) central ion. Consequently, the nuclear spin states contained in the [TbPc$_2$]$^0$ can be described as an effective two-qubit system, also known as qu*d*it ($d = 4$, representing the dimension of the qu*d*it).[9-12] The inherent multilevel characteristics, as well as the shielded nature of the nuclear spins against decoherent environmental fluctuations (electronic, magnetic, etc.), have ultimately led to the implementation of Grover's quantum algorithm on a single [TbPc$_2$]$^0$ molecule.[9d,10]

Two main characteristics allow the use of the [TbPc$_2$]$^0$ molecule as qu*d*it: (*i*) the inherent multilevel properties (qu*d*its where $d > 2$) and (*ii*) the presence of *hf*-QTM events. Due to the multiplicity in [TbPc$_2$]$^0$, entanglement and superposition of multiple states can be achieved in qu*d*its in large dimensions with smaller clusters of processing units;[9,12] whilst the resonant *hf*-QTM allows the manipulation and read-out of the nuclear spin states *via* the reversal of the electronic spins.[9d,e]

In this work, we show how the ferromagnetic interaction between the electronic spins in a dinuclear complex, namely [Tb$_2$Pc$^{Hx8}$Pc$_2$] (**1**), increases the multiplicity of nuclear spins states available for manipulation. Our results show for the first time [despite a large number of previously reported triple decker lanthanide complexes][13], that the reversal of the electronic spins occurs *via* cooperative tunneling (co-tunneling) at specific level crossings, induced by the interaction operating between the electronic states of the Tb(III) ions.

**RESULTS AND DISCUSSION**

***Synthesis and crystal structure***: The studied SMM, namely [Tb$_2$Pc$^{Hx8}$Pc$_2$] (where Pc = phthalocyaninato and $^{Hx8}$Pc = 2,3,9,10,16,17,23,24-octahexylphthalocyaninato) (**1**), was synthesized following a modified reported procedure (see SI for details).[14] The complex is extremely robust, allowing its purification *via* column chromatography. **1** comprises an asymmetric terbium dinuclear complex, composed of two Pc and one $^{Hx8}$Pc ligands (Fig. 1). The complex crystallizes in the monoclinic $P2_1/c$ space group with four molecules residing in the unit cell. Two molecules are related by an inversion center with the other two generated by screw and glide plane symmetries. Coincidentally, both sets of molecules are nearly parallel, with a small tilting angle of 6º between them. At the metal site, each Tb(III) has a distorted square antiprism local symmetry: Tb(1) is sandwiched between the two Pc ligands with a CShM[15] value of 1.593, while Tb(2) is sandwiched between one Pc and one $^{Hx8}$Pc (CShM of 2.553) (See Table S2). Both Tb(III) ions possess a more distorted coordination environment than other members of the parent $[TbPc_2]^{0,-1}$ family.[3d] The intramolecular Tb···Tb distance observed in the crystal structure is 3.5230(8) Å, while the shortest intermolecular Tb···Tb distance is 10.8571(7) Å, thus intermolecular interactions are expectedly small.

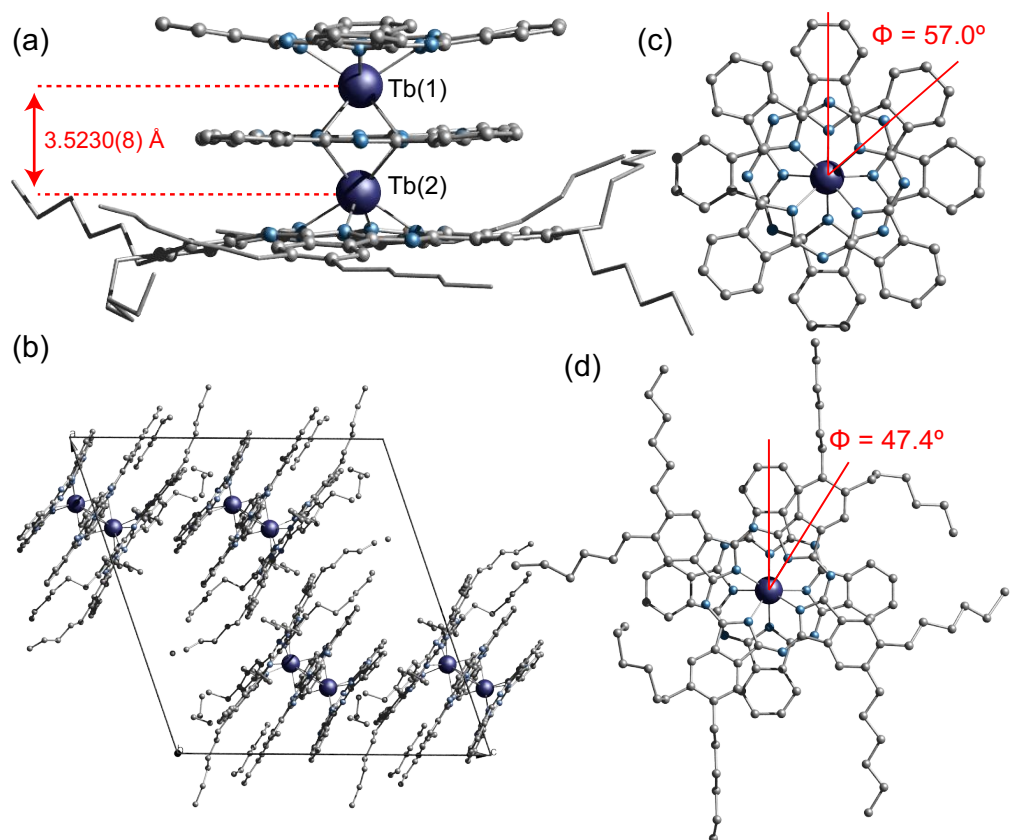

**Figure 1.** (a) Side view of the crystal structure of $[Tb_2Pc^{Hx8}Pc_2]$ (**1**). The red double arrow indicates the Tb···Tb intramolecular distance. (b) Representation of the packing diagram showing a quasi-single orientation of the $[Tb_2Pc^{Hx8}Pc_2]$ units within the crystal lattice (view along the [010] plane). (c) and (d) show the respective skew angles for the Tb(III) sandwiched between two Pc groups and one Pc and $^{Hx8}$Pc groups. Color code: Tb, dark blue; N, cyan; C, gray. H atoms omitted for clarity.

*Magnetic description:* Magnetic measurements were conducted employing the neat sample **1** and a magnetically diluted sample, i.e. 1% [Tb$_2$Pc$^{Hx8}$Pc$_2$] into 99% [Y$_2$Pc$^{Hx8}$Pc$_2$] (**1**$^{dil}$), where intermolecular dipolar fields are strongly reduced. The investigations were conducted in the region of 2 K ≤ $T$ ≤ 300 K using a commercial SQUID magnetometer. The temperature dependent magnetic susceptibility $\chi_M T(T)$ of the powder sample **1** exhibits a room temperature value in agreement with the expected value for two non-interacting Tb(III) ions, *i.e.* 23.5 cm$^3$ mol$^{-1}$ K, *cf.* 23.6 cm$^3$ mol$^{-1}$ K (for two Tb(III) with $g_J$ = 3/2; $J$ = 6). Upon cooling $\chi_M T(T)$ stays practically constant up to ca. 16 K, where it sharply increases to 35.3 cm$^3$ mol$^{-1}$ K due to ferromagnetic interactions between the Tb(III) ions, caused probably by a combination of dipolar and exchange interactions (*vide infra*).[13-14] The same behavior is observed in **1**$^{dil}$ (Fig. 2 and S3). Likewise, magnetization (*M*) *vs.* applied field (*H*) studies between 2 to 5 K (from 0 to 7 T) show that the saturation value for **1** is reached at relatively low fields (ca. 1 T), leading to an *M*(*H*) value of 9.2 µ$_B$ at 7 T. Moreover, alternating magnetic susceptibility studies, conducted at zero DC field for both samples **1** and **1**$^{dil}$, show a frequency-dependent magnetic behavior characterized by a single relaxation process. [Tb$_2$Pc$^{Hx8}$Pc$_2$] (**1**) can, therefore, be described as an SMM (Fig. 2b and S4-5).

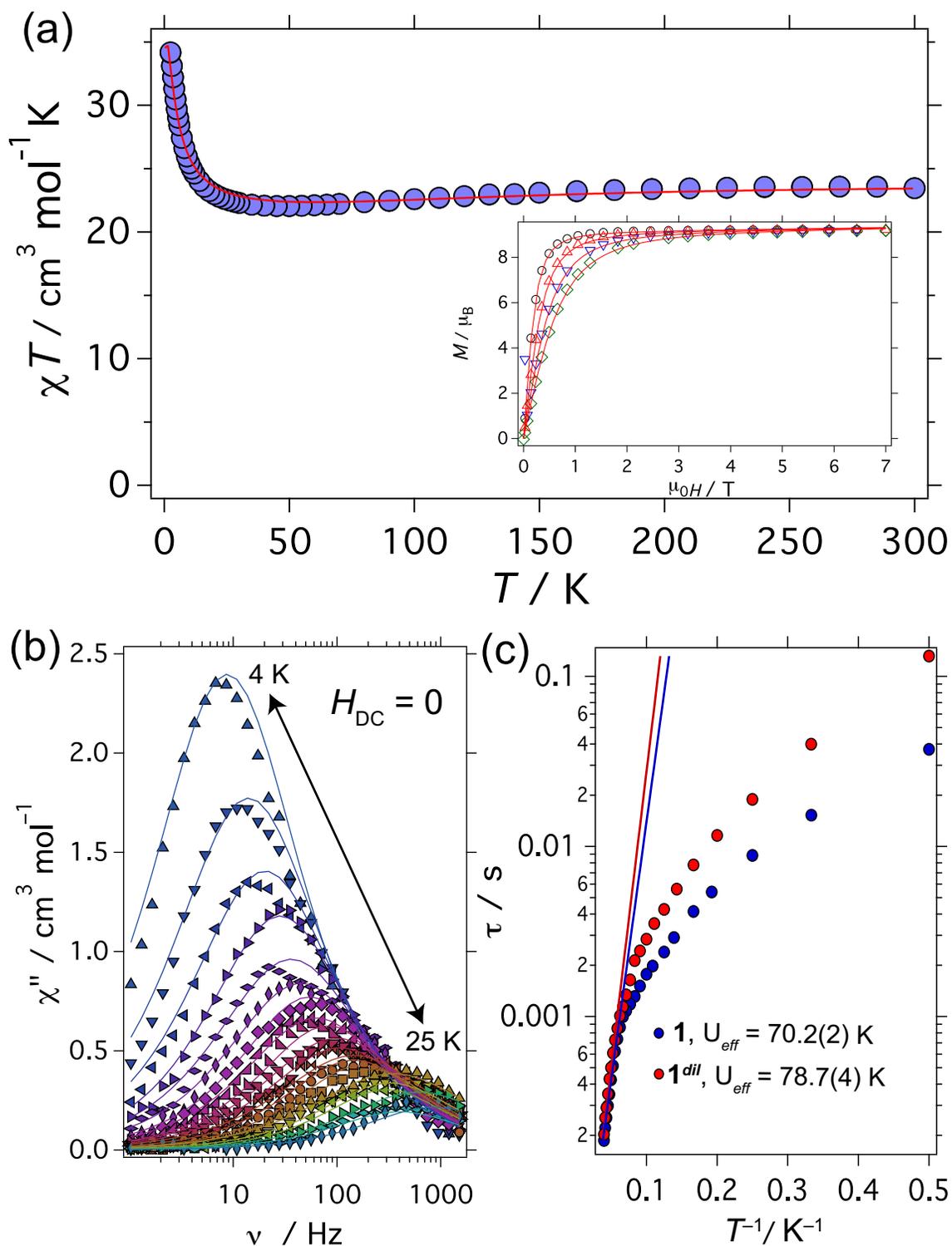

**Figure 2.** (a) $\chi_M T(T)$ and $M(H)$ (inset) data for compound **1** and simulation (solid lines) using the Hamiltonian (3) and parameters described in the text. (b) $\chi''(\nu)$ experimental data for **1**$^{dil}$ with $H_{DC} = 0$ and an oscillating field of 3.5 Oe. Solid lines are fits to a single Debye process. (c) $\tau$ vs. $1/T$ data for complex **1** (blue) and **1**$^{dil}$ (red) and Arrhenius analysis (solid lines).

***Single-crystal studies:*** The nuclear spins embodied in the $^{159}$Tb(III) metal ions, the small interaction between the ions, and the SMM character make **1** an excellent candidate for investigating *hf*-QTM. The important question here is whether the observed small ferromagnetic interaction occurring between electronic spins of the Tb(III) allows the coupling of the Tb(III) nuclear spin, thus increasing the number of accessible nuclear states which could be utilized for testing quantum algorithms. In order to answer this question, we studied a single crystal of **1**$^{dil}$ at sub-Kelvin temperatures with the μ-SQUID technique.[16] Employing the transverse field method, the magnetic field was applied along the mean easy axis of magnetization, which lies close to the [101] crystallographic plane.[17] Figure 3 shows open hysteresis loops at different magnetic field sweep rates and temperatures (Fig. 3 and S6) confirming the SMM behavior. In particular, the hysteresis loops in the vicinity of $\mu_0 H_z = 0$ show a staircase-like structure with seven main transitions occurring at 0, ±15.4, ±30.4 and ±45.7 mT (Fig. 3 and 4a,b). Furthermore, two additional broad transitions are observed at ±550 mT (Inset in Fig. 4b and S8). This observation of seven transitions is different but also analogous to the behavior of the parent mononuclear [TbPc$_2$]$^{±1,0}$ complex that shows only four *hf*-QTM transitions at ±12 mT and ±37 mT. These correspond to the avoided level crossings that conserve the nuclear spin of the Tb(III) ion. The tunnel splittings are the result of the transverse ligand field interactions which are caused by the transverse crystal field terms ($A_4^4 \hat{O}_4^4$ and $A_6^4 \hat{O}_6^4$) arising from small distortion of $D_{4d}$ symmetry of the molecule.[3d] Note that a recently reported fused Pc-bridged Tb-dimer, with a relatively long intramolecular Tb···Tb distance of 11.3135(7) Å exhibits only four *hf*-QTM transitions,[18] suggesting that the two Tb-containing moieties act as two rather independent SMM-units. Furthermore, a mixed heteronuclear Tb(III)-Dy(III) dimeric SMM with almost identical geometrical parameters as found in **1** does show only four *hf*-QTM transitions.[14b] The lack of additional QTM transitions in the dimeric Tb(III)-Dy(III) SMM could be ascribed to the isotopic mixture of nuclear states in the Dy(III) ions as well as smaller or quenched Tb(III)···Dy(III) interactions.

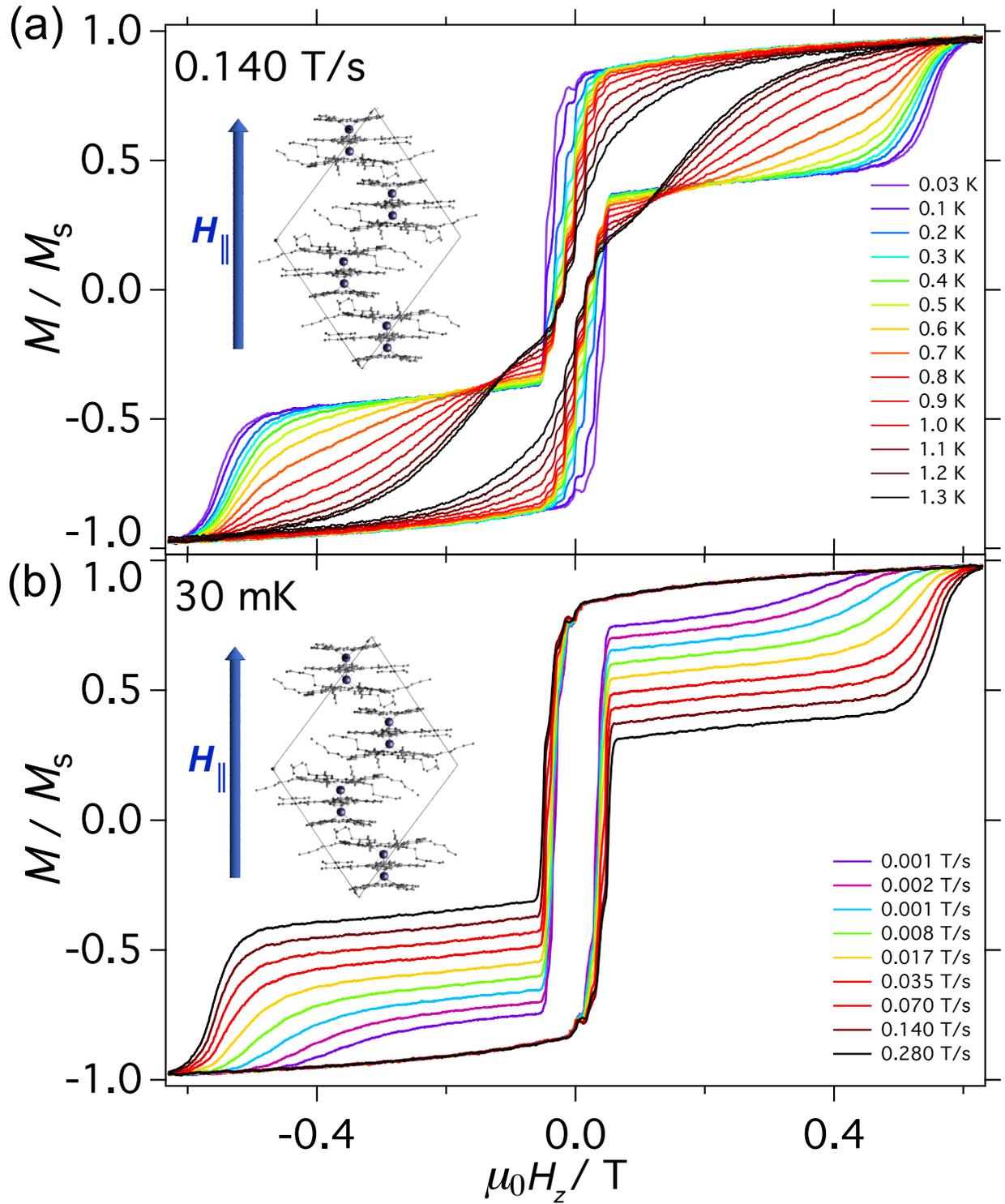

**Figure 3.** (a) Temperature dependence of the magnetization of **1**$^{dil}$ at a field sweep rate of 0.140 T/s and (b) Field dependence of the magnetization at $T$ = 30 mK. The field was applied parallel to the easy axis of the magnetization. Before each field sweep, a waiting time of more than 1000 s at ±1 T was used to thermally equilibrate the nuclear spin system with the thermal bath.

In the following, it will be shown that the electronic states of the Tb(III) ions of [Tb$_2$Pc$^{Hx8}$Pc$_2$] are effectively coupled and the observed multitude of *hf*-QTM steps can be assigned to a co-tunneling of the electronic spin $|J_z = \pm 6\rangle$, while the nuclear spin states are conserved. For this purpose, we choose a ligand field (lf) Hamiltonian that takes into account the distorted electronic environment of the two Tb(III) ions in **1**$^{dil}$, imposing locally an approximate $C_4$ symmetry[3d,20,21]:

$$\mathcal{H}_{lf}^i = \alpha A_2^0 \hat{O}_2^0 + \beta\left(A_4^0 \hat{O}_4^0 + A_4^4 \hat{O}_4^4\right) + \gamma\left(A_6^0 \hat{O}_6^0 + A_6^4 \hat{O}_6^4\right) \quad (1)$$

where $i$ = 1 or 2 refers to each Tb(III) ion), α, β and γ are the Stevens coefficients, $\hat{O}_k^q$ are the equivalent Stevens operators, and $A_k^q$ are the ligand field parameters. In order to account for the effect of the magnetic field and the Tb(III) nuclear spin on the multiplicity of the $m_J$ states, three other terms are included:

$$\mathcal{H}_{Tb}^i = \mathcal{H}_{lf}^i + g_J\mu_0\mu_B \mathbf{J}_z^i \cdot \mathbf{H}_z + A_{hf}\mathbf{I}^i\cdot\mathbf{J}_z^i + P\left((I_z^i)^2 - \frac{1}{3}(I+1)I\right) \quad (2)$$

in (2) the second term describes the Zeeman interaction and the third and fourth term describe the hyperfine and the quadrupole interaction respectively, with $A_{hf}$ and $P$ the corresponding parameters. The ligand field parameters of a closely related molecule have been experimentally determined from NMR and magnetic susceptibility data, where the $A_k^q$ parameters are slightly different for each Tb(III)-site reflecting the asymmetry of the molecule.[20] Likewise, the $A_4^4$ and $A_6^4$ terms in ref. 20 are larger than that of the mononuclear [TbPc$_2$]$^-$ analogues.[3d] Despite the differences of the $A_k^q$, compared to the parameters of the archetypal [TbPc$_2$]$^-$ complex, the ground state for each Tb(III) is well defined as $|J_z = \pm 6\rangle$. Given that the accurate determination of high order terms is rather difficult, in our analysis we solely employed the $A_k^q$ (where $k$ = 2, and 4, whilst $q$ = 0) terms reported in ref. 20, i.e. $A_2^0$ = +289 cm$^{-1}$ and $A_4^0$ = –209 cm$^{-1}$ for Tb(1); and $A_2^0$ = +293 cm$^{-1}$, $A_4^0$ = –197 cm$^{-1}$ for Tb(2), whilst the remaining terms are set to zero. Due to the close proximity of the Tb(III) ions, i.e. an intramolecular Tb···Tb distance 3.5230(8) Å (*vide supra*), both ions are connected by a weak dipolar interaction of the form[13c] ($\mathcal{H}_{dip}$) (See section 3 of supplementary information for a more detailed description). Thus, the Hamiltonian for **1**$^{dil}$ reads:

$$\mathcal{H} = -2J_1 \cdot \mathcal{H}_{dip} \cdot J_2 + \mathcal{H}_{Tb}^1 + \mathcal{H}_{Tb}^2 \quad (3)$$

The energy diagram of **1**$^{dil}$ can be calculated by exact diagonalization of the $(2J+1)^2(2I+1)^2 \times (2J+1)^2(2I+1)^2$ Hamiltonian (3). For simplicity, we assume $P$ and $A_{hf}$ for both sites to be equal. Employing $A_{hf}$ and $P$ as parameters, we are able to reproduce the seven QTM events observed in the μ-SQUID data with $P = +0.010$ cm$^{-1}$ and $A_{hf} = +0.0215$ cm$^{-1}$ (Fig. 4c) provided that the $\mathcal{H}_{dip}$ is larger than the hyperfine coupling (*vide infra*). We find a $P$ parameter equal to the one found for [TbPc$_2$]$^-$, whilst a slightly larger $A_{hf}$ than in the mononuclear case is obtained (cf. $P = +0.010$ cm$^{-1}$ and $A_{hf} = +0.0173$ cm$^{-1}$ for [TbPc$_2$]$^-$).[3d] Note that the effect of the 6º tilting angle between the two differently oriented molecules in the unit cell causes only a small broadening of each crossing point and does not account for additional QTM events.

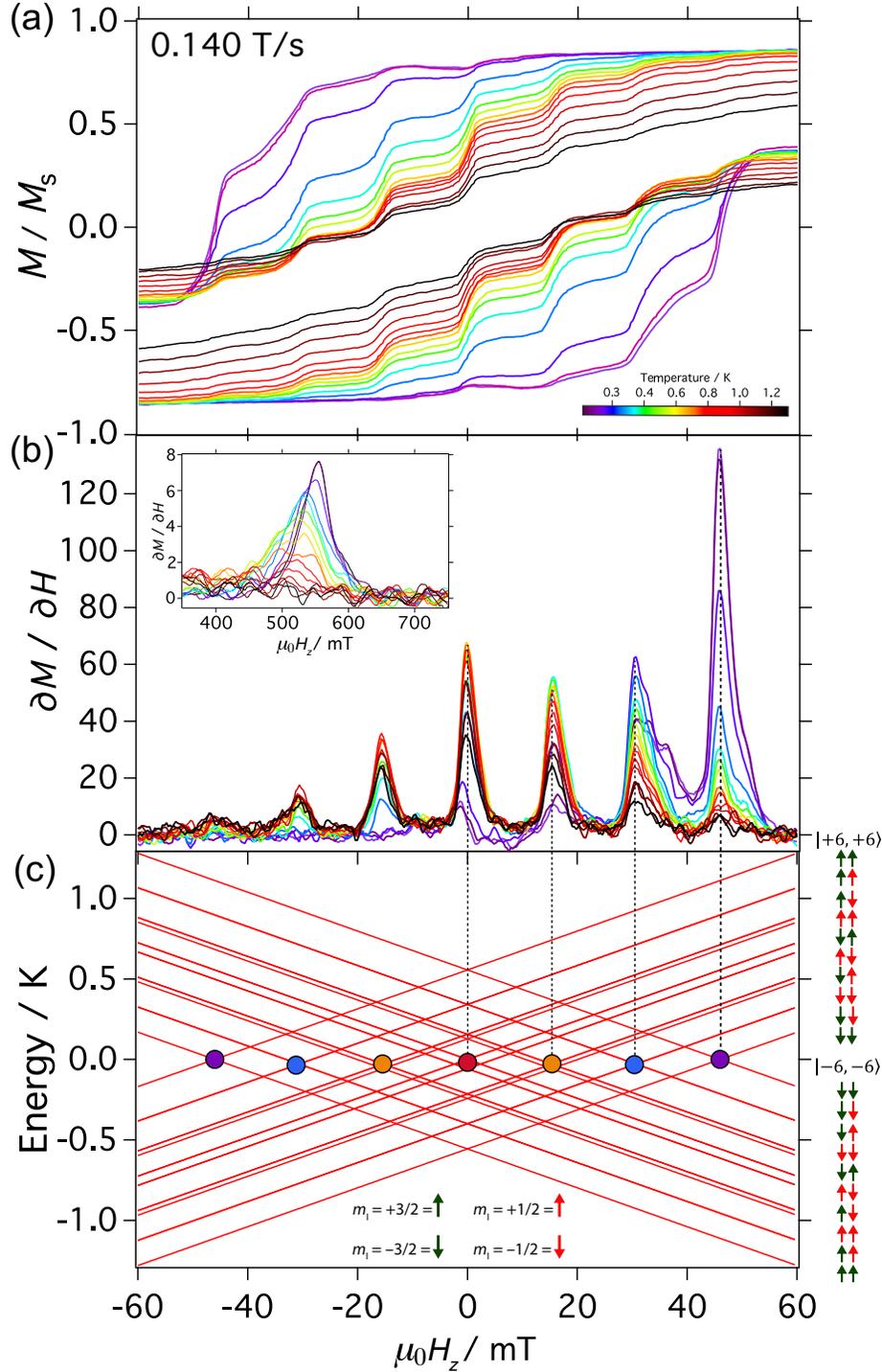

**Figure 4.** (a) Zoom of the hysteresis loops of Fig. 3a recorded between ±1 T with a field sweep of 0.140 T/s at several temperatures, showing a staircase-like structure. (b) First field derivative for a field sweep from −1 T to +1 T of the data in Fig. 4a. (c) Simulated Zeeman diagram with the field parallel to the easy axes, employing (3) and $A_k^q$ (where $k$ = 2, and 4, whilst $q$ = 0) described in the text. *hf*-QTM are ascribed to co-tunneling of the electronic spins, that is, transitions between $|-6, I_z^a\rangle|-6, I_z^b\rangle \leftrightarrow |+6, I_z^a\rangle|+6, I_z^b\rangle$. The arrows to the rightmost side are the pictorial representation of the nuclear spins for every $m_I$ state (See Fig. S7 for a detailed description of all states).

Numerical diagonalization of the Hamiltonian given by equation (3) results in 100 level crossings close to zero field and for the $|J_z = \pm 6\rangle|J_z = \pm 6\rangle$ states (Fig. 4c). However, out of these only 10 have large tunnel splittings leading to a high probability of electronic spin reversal [only seven are observed, as for some crossings tunneling occur at the same resonance field]. The tunnel splittings are induced by off-diagonal terms in the Hamiltonian, which couple the $|J_z = \pm 6\rangle$ states. In the terbium dimer, both $^{159}$Tb(III) ions are coupled by ferromagnetic interaction (*vide infra*), and the eigenstates can be expressed as $|J_z^a, I_z^a\rangle|J_z^b, I_z^b\rangle$. At very low temperatures, solely the $|-6, I_z^a\rangle|-6, I_z^b\rangle$ states are populated (for positive fields). Before each field sweep the nuclear spins are thermalized. For this, a waiting time of more than 1000 s at ±1 T was used to thermally equilibrate the nuclear spin system with the thermal bath. The electronic spin reversal occurs at level crossings with large tunnel splittings when sweeping the field from a high negative field to a positive one.

The seven events observed in Fig. 4b can be rationalized as follows: upon sweeping the field between ±1 T, tunneling occurs while conserving the nuclear states. At zero field, two tunneling events occur between the $\left|-6, \pm\frac{1}{2}\right\rangle\left|-6, \mp\frac{1}{2}\right\rangle$ to $\left|+6, \pm\frac{1}{2}\right\rangle\left|+6, \mp\frac{1}{2}\right\rangle$ and also between the $\left|-6, \pm\frac{3}{2}\right\rangle\left|-6, \mp\frac{3}{2}\right\rangle$ to $\left|+6, \pm\frac{3}{2}\right\rangle\left|+6, \mp\frac{3}{2}\right\rangle$ states (red circle in Fig. 4c). Two other allowed tunneling events occur at ±15.4 mT, where the reversal occurs *via* the states $\left|-6, \pm\frac{1}{2}\right\rangle\left|-6, \pm\frac{1}{2}\right\rangle$ to $\left|+6, \pm\frac{1}{2}\right\rangle\left|+6, \pm\frac{1}{2}\right\rangle$ and $\left|-6, \pm\frac{1}{2}\right\rangle\left|-6, \mp\frac{3}{2}\right\rangle$ to $\left|+6, \pm\frac{1}{2}\right\rangle\left|+6, \mp\frac{3}{2}\right\rangle$ (orange circles in Fig. 3c). At ±30.4 mT the reversal is permitted *via* $\left|-6, \pm\frac{1}{2}\right\rangle\left|-6, \pm\frac{3}{2}\right\rangle$ to $\left|+6, \pm\frac{1}{2}\right\rangle\left|+6, \pm\frac{3}{2}\right\rangle$ (blue circles in Fig. 4c), whilst the last event at ±45.7 mT, is ascribed to the electronic spins flip *via* $\left|-6, \pm\frac{3}{2}\right\rangle\left|-6, \pm\frac{3}{2}\right\rangle$ to $\left|+6, \pm\frac{3}{2}\right\rangle\left|+6, \pm\frac{3}{2}\right\rangle$ states (purple circles in Fig. 4c) [See Fig. S7 for a detailed description of all states]. Therefore, we conclude that simultaneous spin reversals at specific avoided crossings are observed, as long as the nuclear spin states are strictly conserved. As observed in Fig. 4b, the co-tunnel probability not only depends on the tunnel splittings but also on the thermal population of the levels, yielding a strong temperature dependence of step height of each transition.

Additionally, at $\mu_0H_z = \pm550$ mT, two broader transitions are observed (inset in Fig. 4b), which are in agreement with a ferromagnetic interaction between the Tb(III) ions, where single spin flips occur between the electronic ground states, *i.e.* $|\pm J_z^a, I_z^a\rangle|\mp J_z^b, I_z^b\rangle \leftrightarrow |\pm J_z^a, I_z^a\rangle|\pm J_z^b, I_z^b\rangle$ or $|\mp J_z^a, I_z^a\rangle|\pm J_z^b, I_z^b\rangle \leftrightarrow |\mp J_z^a, I_z^a\rangle|\mp J_z^b, I_z^b\rangle$ (See Fig. S8).[13c] The hyperfine structure is not experimentally resolved, probably due to small distributions of ligand field parameters and misalignment. As can be seen in Fig. S8, these transitions at $\mu_0H_z = \pm550$ mT cannot be reproduced employing a purely dipolar coupling between the Ising $|J_z = \pm6\rangle$ states, which places the first excited state at +3.21 cm$^{-1}$ from the ground state (calculated for a Tb⋯Tb distance of 3.5230(8) Å, using a point-dipole approximation). Experimentally, the first excited state lies at about +4.6 cm$^{-1}$, which is larger than the calculated dipolar value, therefore it is possible that, in addition, a small contribution of exchange interaction is present. To assess this possibility, we add an isotropic Heisenberg interaction ($J_{ex}$) to the dipolar matrix. The Hamiltonian has the following form:

$$\mathcal{H} = -2J_1 \cdot (\mathcal{H}_{dip} + J_{ex}) \cdot J_2 + \mathcal{H}_{Tb}^1 + \mathcal{H}_{Tb}^2 \quad (4)$$

The simulations were adjusted so that the crossing point between the ground state and the first excited state observed through μ-SQUID data at ±500 mT is reproduced. Employing (4) we are able to reproduce the crossing point between the first excited state and the ground state with $J_{ex}$ = +0.0097 cm$^{-1}$. Note that solely from the magnetic data it is impossible to accurately assess the possibility of exchange interactions occurring in [Tb$_2$Pc$^{Hx8}$Pc$_2$] given that both $\mathcal{H}_{dip}$ and ($\mathcal{H}_{dip} + J_{ex}$) equally reproduce the data. Our results show that the possibility of exchange interactions occurring between the Tb(III) is highly probable and requires further studies.

**CONCLUSIONS**

Hyperfine driven quantum tunneling of the magnetization has been previously observed in several lanthanide complexes[3d,19], however, what makes [Tb$_2$Pc$^{Hx8}$Pc$_2$] really unique is the collective behavior triggered by the small interaction between the two Tb(III) ions. This interaction leads to a coupling of the nuclear spins of the two $^{159}$Tb(III) metal ions, causing not just *hf*-QTM occurrences, as observed in the [TbPc$_2$]$^-$, but additionally increasing the multiplicity of nuclear spin states and, in consequence, allowing the *hf*-QTM at additional resonance field positions. Moreover, the results reported here differ from that of the exchange bias QTM, where the resonance fields are shifted due to exchange with adjacent nuclei.[3]

In conclusion, the resonant QTM has been investigated in a dimeric Tb(III)-based SMM *via* μ-SQUID measurements, allowing the determination of the hyperfine and quadrupolar parameters. The *hf*-QTM events are ascribed to the simultaneous reversal (co-tunneling) of the electronic spin, while the nuclear spins are conserved. As in the [TbPc$_2$]$^{\pm 1,0}$ case, the QTM transitions corresponding to each nuclear spin state could be used to the read-out of the nuclear spins in [Tb$_2$Pc$^{Hx8}$Pc$_2$] (**1**) in a qu*d*it scheme as enlarged quantum register. Note that in the archetypal [TbPc$_2$]$^{\pm 1,0}$, just four states are accessible (corresponding to the $m_I = \pm 3/2$ and $\pm 1/2$ states)[9] limiting the applicability of multilevel quantum algorithms. However, in **1**, the observation of enlarged multiplicity of states by cooperative electronic coupling would open a general avenue to the creation of larger quantum directories, thus allowing the realization of a molecular spin qu*d*it with a exploitable dimension of $d = 16$ (with $d = (2I+1)^n$ where $n = 2$; $I = 3/2$).[10]

## AUTHOR INFORMATION

### Corresponding Authors

* Mario.ruben@kit.edu; eufemio.pineda@kit.edu; wolfgang.wernsdorfer@kit.edu

### Notes

The authors declare no competing financial interest.

## ACKNOWLEDGEMENTS

We acknowledge the DFG-TR 88 "3Met" (project A8) and the Karlsruhe Nano Micro Facility (KNMF, www.kit.edu/knmf) for the provision of access to instruments at their laboratories. WW thanks the A. v. Humboldt foundation and the ERC grant MoQuOS Nr. 741276.

**Supporting Information Available**: cif files, further synthetic details, structural and magnetic plots.

entanglement between molecular qubits with four-dimensional inelastic neutron scattering. *Nat. Commun.* **2016**, *8*, 14543.

7. Troiani, F.; Affronte, M. Molecular spins for quantum information technologies. *Chem. Soc. Rev.* **2011**, *40*, 3119. (b) Lehmann, J.; Gaita-Ariño, A.; Coronado, E.; Loss, D. Molecular spintronics and quantum computing. *J. Mater. Chem.* **2009**, *19*, 1672–1677.

8. Ishikawa, N.; Sugita, M.; Ishikawa, T.; Koshihara, S.-Y.; Kaizu, Y. Lanthanide Double-Decker Complexes Functioning as Magnets at the Single-Molecular Level. *J. Am. Chem. Soc.* **2003**, *125,* 8694–8695.

9. (a) Vincent, R.; Klyatskaya, S.; Ruben, M.; Wernsdorfer, W.; Balestro, F. Electronic read-out of a single nuclear spin using a molecular spin transistor. *Nature*, **2012**, *488*, 357–360; (b) Thiele, S.; Balestro, F.; Ballou, R.; Klyatskaya, S.; Ruben, M.; Wernsdorfer, W. Electrically driven nuclear spin resonance in single-molecule magnets. *Science* **2014**, *344*, 1135–1138; (c) Godfrin, C.; Thiele, S.; Ferhat, A.; Klyatskaya, S.; Ruben, M.; Wernsdorfer, W.; Balestro, F. Electrical Read-Out of a Single Spin Using an Exchange-Coupled Quantum Dot. *ACS Nano*, **2017**, *11*, 3984–3989; (d) Godfrin, C.; Ferhat, A.; Ballou, R.; Klyatskaya, S.; Ruben, M.; Wernsdorfer, W.; Balestro, F. Operating Quantum States in Single Magnetic Molecules: Implementation of Grover's Quantum Algorithm. *Phys. Rev. Lett.* **2017**, *119*, 187702; (e) Moreno-Pineda, E.; Godfrin, C.; Balestro, F.; Wernsdorfer, W.; Ruben, M. Molecular spin qudits for quantum algorithms. *Chem. Soc. Rev.* **2018**, *47*,501–513.

10. Morello, A. Quantum search on a single-atom qudit. *Nat. Nanotech.* **2018**, *13*, 9-10.

11. (a) Jenkins, M. D.; Duan, Y.; Diosdado, B.; García-Ripoll, J. J.; Gaita-Ariño, A.; Giménez-Saiz, C.; Alonso, P. J.; Coronado, E.; Luis, F. Coherent manipulation of three-qubit states in a molecular single-ion magnet. *Phys. Rev. B*, **2017**, *95*, 064423; (b) Martínez-Pérez, M. J.; Cardona-Serra, S.; Schlegel, C.; Moro, F.; Alonso, P. J.; Prima-García, H.; Clemente-Juan, J. M.; Evangelisti, M.; Gaita-Ariño, A.; Sesé, J.; van Slageren, J.; Coronado, E.; Luis, F. Gd-Based Single-Ion Magnets with Tunable Magnetic Anisotropy: Molecular Design of Spin Qubits. *Phys. Rev. Lett.* **2012**, *108*, 247213.

12. (a) O'Leary, D. P.; Brennen, G. K.; Bullock, S. S. Parallelism for quantum computation with qudits. *Phys. Rev. A* **2006**, *74*, 032334; (b) Smith, A.; Anderson, B. E.; Sosa-Martinez,

**For Table of Contents Only**

Magnetic studies on a dimeric terbium complex show that the coupling between the Tb(III) metal ions induces a multitude of hyperfine-driven quantum tunneling events. As a result, the [Tb$_2$Pc$^{Hx8}$Pc$_2$] complex represents a qu*d*it (with $d = 16$), where the coupling of the nuclear spin states via the electronic spins is observed for the first time. This result may allow the molecule to operate for quantum gates in the implementation of quantum algorithms.

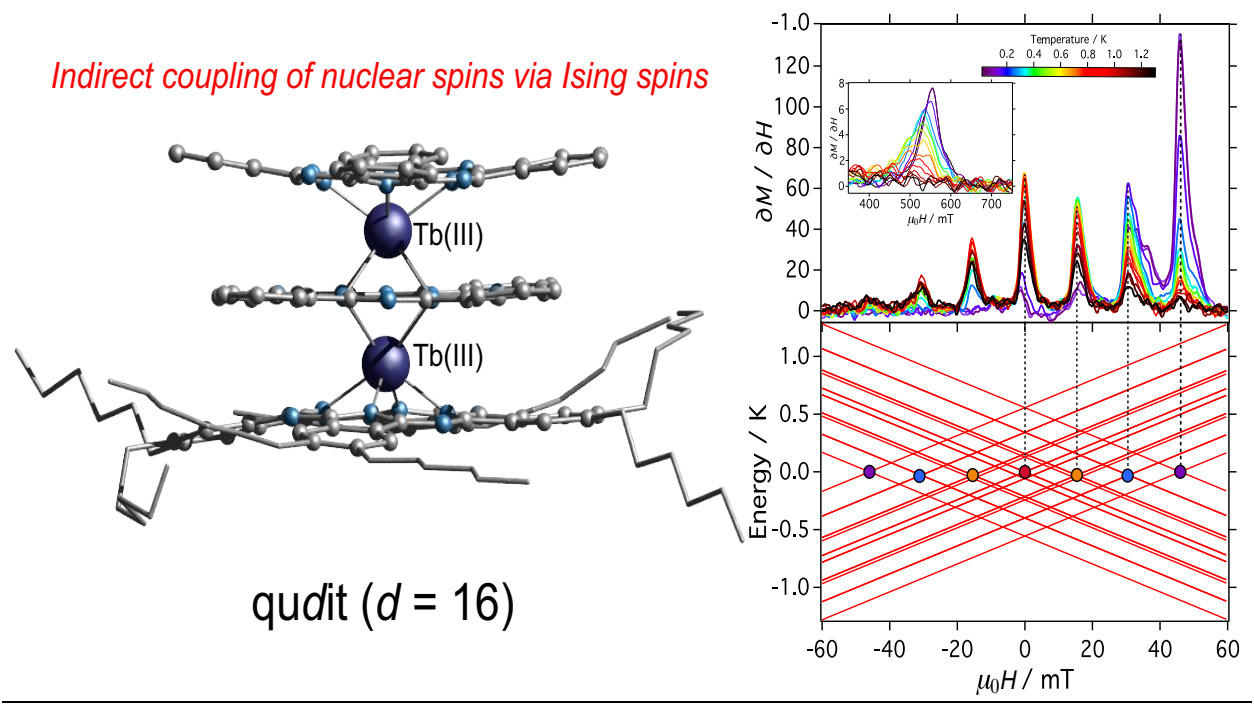

# Supporting Information

## *Observation of Cooperative Electronic Quantum Tunneling: Increasing Accessible Nuclear States in a Molecular Qudit*

Eufemio Moreno-Pineda[a,]*, Svetlana Klyatskaya[a], Ping Du[a], Marko Damjanović[a], Gheorghe Taran[b,c], Wolfgang Wernsdorfer[a,b,c]* and Mario Ruben [a,d,]*

  a. *Institute of Nanotechnology (INT), Karlsruhe Institute of Technology (KIT), Hermann-von-Helmholtz-Platz 1, D-76344 Eggenstein-Leopoldshafen, Germany. E-mail: eufemio.pineda@kit.edu; wolfgang.wernsdorfer@kit.edu; mario.ruben@kit.edu*
  b. *Physikalisches Institut, Karlsruhe Institute of Technology, D-76131 Karlsruhe, Germany*
  c. *Institut Néel, CNRS / Université Grenoble Alpes, 25 rue des Martyrs, F-38000 Grenoble, France.*
  d. *Institut de Physique et Chimie des Matériaux de Strasbourg (IPCMS), CNRS-Université de Strasbourg, 23 rue du Loess, BP 43, F-67034 Strasbourg Cedex 2, France.*

**Experimental Details**

Reactions sensitive to oxygen and moisture were conducted under an argon atmosphere. The glassware was oven-dried at 140°C. All reagents were purchased from commercial sources and used as received. Phthalocyanine with eight hexyl groups ($^{Hx8}$Pc) (**4**) and lanthanide bis-phthalocyanines LnPc$_2$ (where Ln = Tb(III) (**5**) and Y(III) (**6**)) were prepared according to the literature procedures.[1]

**A. Synthetic Methods**

The homonuclear complexes PcLnPcLn$^{Hx8}$Pc (Ln= Tb (**1**), Y (**2**), Scheme 1) were synthesized by the in-situ fusing of the free base Pc ligand with 2,3,9,10,16,17,23,24-octahexyl substituents, $^{Hx8}$PcH$_2$ (**4**)[1a] and bis-phthalocyanine complexes [LnPc$_2$]$^0$ (Ln= Tb (**4**), Y (**5**))[1b] with excess of corresponding Ln(acac)$_3$·$n$H$_2$O(**7**, **8**) in 1,2,4-Cl$_3$-benzene affording the targeted crude product. Pure complexes (**1**) and (**2**) as a dark green solid were separated from the crude mixtures by column chromatography (basic alumina oxide) with an eluent of CH$_2$Cl$_2$/CHCl$_3$ (20:1). The complexes (**1**) and (**2**) are composed of three Pc ligands and two Ln$^{3+}$ ions, resulting in a neutral complex with a closed shell p-electron system.

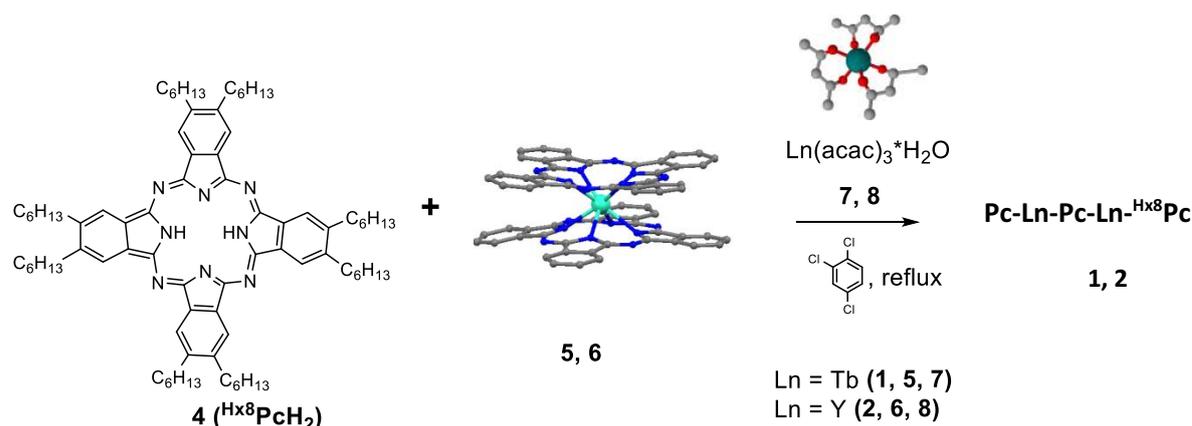

**Scheme S1.** Synthesis of [Pc-Ln-Pc-Ln-$^{Hx8}$Pc] complexes, Ln = Tb (**1**), Ln = Y (**2**) by thermal fusing.

**General procedure for the synthesis of [Tb$_2$Pc$_2$$^{Hx8}$Pc] (1) and [Y$_2$Pc$_2$$^{Hx8}$Pc] (2)**: Under a slow stream of Ar, a mixture of 0.042 mmol of free base $^{Hx8}$PcH$_2$ (**4**), 0.043 mmol of LnPc$_2$ (**5, 6**) and 0.175 mmol of Ln(acac)$_3$ (**7, 8**) were thermally fused in 5 mL of 1,2,4-Cl$_3$-benzene. The mixture was heated on an oil bath set at 60 °C with mechanical stirring until the entire solid was dissolved, then the bath temperature was increased, and the mixture was refluxed for 1 h with careful monitoring by TLC/UV. The reaction mixture was cooled to room temperature, chromatographed on a Al$_2$O$_3$ column (basic, grade IV). Using a dichloromethane/chloroform mixture (20/1) as an eluent, a blue-green band of the aimed substance was collected. The solution was concentrated, methanol was added, and the formed precipitate was collected by filtration.

The final product PcLnPcLn$^{Hx8}$Pc (**1, 2**) was obtained as a dark blue powder. The compounds were identified by MALDI-ToF.

**[Tb$_2$Pc$_2$$^{Hx8}$Pc]** (**1**), 76 mg (72 %). MALDI-ToF calculated for C$_{144}$H$_{144}$N$_{24}$Tb$_2$ [MH]$^+$: 2528.0542. Found 2528.0881 (M$^+$, 100 %); Elemental analysis (%) calcd. for calculated for C$_{144}$H$_{144}$N$_{24}$Tb$_2$: C 68.40, N 13.29, H 5.74: found: C 68.33, N 13.27, H 5.82; **[Y$_2$Pc$_2$$^{Hx8}$Pc]** (**2**), 69 mg (69 %) mg (71). MALDI-ToF calculated for C$_{144}$H$_{144}$N$_{24}$Y$_2$ [MH]$^+$: 2388.0122. Found 2388.0396 (M$^+$, 100 %); C$_{144}$H$_{144}$N$_{24}$Y$_2$: C 72.41, N 14.07, H 7.44: found: C 72.51, N 13.98, H 7.42.

**The diluted complex (1$^{dil}$):** The magnetically dilute sample, i.e. 1% [Tb$_2$Pc$^{Hx8}$Pc$_2$] into 99% [Y$_2$Pc$^{Hx8}$Pc$_2$] (**1$^{dil}$**), was obtained by combining accurately measured amounts of [Tb$_2$Pc$_2$$^{Hx8}$Pc] (**1**) and [Y$_2$Pc$_2$$^{Hx8}$Pc] in 1:99 molar ratio. The two compounds were solubilized in hot hexane and recrystallized by slow evaporation affording very regular dark blue crystals.

**B. Crystallography**

Single crystal X-ray diffraction data of **1** and **2** were collected employing an STOE StadiVari 25 diffractometer with a Pilatus300 K detector using GeniX 3D HF micro focus with MoKα radiation (λ = 0.71073 Å). The structure was solved using direct methods and was refined by full-matrix least-squares methods on all $F^2$ using SHELX-2014[2] implemented in Olex2[3]. The crystals were mounted on a glass tip using crystallographic oil and placed in a cryostream. Data were collected using φ and ω scans chosen to give a complete asymmetric unit. All non-hydrogen atoms were refined anisotropically. Hydrogen atoms were calculated geometrically riding on their parent atoms. Full crystallographic details can be found in CIF format: see the Cambridge Crystallographic Data Centre database (CCDC 1567148–1567149).

**C. Magnetic Measurements**

AC and DC magnetic susceptibility measurements were collected using Quantum Design MPMS-XL SQUID and a VSM SQUID magnetometer on polycrystalline material in the temperature range 2 – 10 K, field ranging from 0 to 7 T and in the frequency range of 1–1500 Hz with an oscillating magnetic field of 3.5 Oe. Low temperature (0.03 – 2 K) magnetization measurements were performed on single crystals using a μ-SQUID apparatus at different sweep rates between 0.002 and 0.280 T s$^{-1}$. The time resolution is approximately 1 ms. The magnetic field can be applied in any direction of the μ-SQUID plane with a precision better than 0.1° by separately driving three orthogonal coils. The magnetic field was applied parallel to the easy axis of magnetization by using the transverse field method.[4] In order to ensure good thermalisation, each sample was fixed with apiezon grease.

# D. Supplementary Tables

## 1. Crystallographic Tables

**Table S1.** Crystallographic information for compounds **1** and **2**.

|  | 1 | 2 |
|---|---|---|
| formula | $C_{150}N_{24}Tb_2H_{158}$ | $C_{150}N_{24}Y_2H_{158}$ |
| FW / g mol$^{-1}$ | 2614.83 | 2474.81 |
| crystal system | monoclinic | monoclinic |
| space group | *P21/c* | *P21/c* |
| *a*/Å | 26.4307(6) | 26.4389(6) |
| *b*/Å | 20.9404(3) | 20.9509(4) |
| *c*/Å | 24.7838(4) | 24.7100(6) |
| *α= γ* /° | 90 | 90 |
| *β* /° | 109.206(2) | 109.153(2) |
| *V*/Å$^3$ | 12953.6(4) | 12929.7(5) |
| *Z* | 4 | 4 |
| *ρ* calcd/g cm$^{-3}$ | 1.341 | 1.271 |
| *T*/K | 180 | 180 |
| μ (Mo K$_α$)/mm$^{-1}$ | 1.145 | 0.956 |
| $R_1(I>2\sigma)(I))^a$ | 0.0654 | 0.0885 |
| $wR_2{}^a$ | 0.1385 | 0.23421 |

$^a$ $R_1 = ||F_0| - |F_c||/|F_0|$, $wR_2=[w(|F_0| - |F_c|)^2/w|F_0|^2]^{1/2}$

2. **SHAPE analysis**

**Table S2.** Continuous shaped measures (CShM) for compound **1** obtained using SHAPE.

| CShM | Tb (1) | Tb (2) |
|---|---|---|
| OP-8 | 36.832 | 37.047 |
| HPY-8 | 23.138 | 23.770 |
| HBPY-8 | 16.713 | 13.078 |
| CU-8 | 9.180 | 5.192 |
| **SAPR-8** | **1.593** | **2.553** |
| TDD-8 | 3.119 | 2.740 |
| JGBF-8 | 18.399 | 17.235 |
| JETBPY-8 | 31.546 | 28.723 |
| JBTPR-8 | 4.043 | 4.930 |
| BTPR-8 | 3.381 | 4.265 |
| JSD-8 | 6.781 | 7.224 |
| TT-8 | 10.031 | 6.090 |
| ETBPY-8 | 25.627 | 25.165 |

OP-8 = ($D_{8h}$) Octagon
HPY-8 = ($C_{7v}$) Heptagonal pyramid
HBPY-8 = ($D_{6h}$) Hexagonal bipyramid
CU-8 = ($O_h$) Cube
SAPR-8 = ($D_{4d}$) Square antiprism
TDD-8= ($D_{2d}$) Triangular dodecahedron
JGBF-8= ($D_{2d}$) Johnson gyrobifastigium J26
JETBPY-8 = ($D_{3h}$) Johnson elongated triangular bipyramid J14
JBTPR-8 = ($C_{2v}$) Biaugmented trigonal prism J50
BTPR-8 = ($C_{2v}$) Biaugmented trigonal prism
JSD-8 = ($D_{2d}$) Snub diphenoid J84
TT-8 = ($T_d$) Triakis tetrahedron
ETBPY-8 = ($D_{3h}$) Elongated trigonal bipyramid

### 3. Dipolar Matrix

Dipolar interactions between two $J = 6$ can be calculated employing the following equation:

$$\mathcal{H}_{dip} = \frac{\mu_B^2}{r^3} - [\bar{\bar{g}}_A \cdot \bar{\bar{g}}_B - 3(\bar{\bar{g}}_A \cdot \vec{R}) \cdot (\vec{R}^T \cdot \bar{\bar{g}}_B)] \quad (S1)$$

Where $\mu_B$ is the Bohr magneton, $r$ is the Tb⋯Tb distance obtained from crystallographic analysis. $\bar{\bar{g}}_{A/B}$ is the g-matrix of ion A and B, and $\vec{R}$ is directional unit vector between the two ions. Due to the highly axial ligand field parameters, the anisotropic axis expectedly lies along the z-axis (perpendicular to the Pc planes) with no component along the x- and y-axis, therefore the the g-matrix for the Tb(III) with $J = 6$ reads:

$$\bar{\bar{g}}_{Tb} = \bar{\bar{g}}_A = \bar{\bar{g}}_B = \begin{pmatrix} 0 & 0 & 0 \\ 0 & 0 & 0 \\ 0 & 0 & 3/2 \end{pmatrix}$$

These two Tb(III) ions are connected by a unit vector of the form:

$$\vec{R} = \begin{pmatrix} 0 \\ 0 \\ 1 \end{pmatrix}$$

Thus, leading to a dipolar matrix for two Tb(III) ions separated by a 3.5230 Å:

$$\mathcal{H}_{dip} = \begin{pmatrix} 0 & 0 & 0 \\ 0 & 0 & 0 \\ 0 & 0 & +0.0223 \end{pmatrix} \text{cm}^{-1} \text{ (the value has been divided by } -1/2 \text{ for a } -2J \text{ Hamiltonian)}$$

This value places the first excited state at +3.2 cm$^{-1}$, which bodes well with the expected value for two $m_J = \pm 6$, estimated by Ishikawa[5] as : $\langle ++|\mathcal{H}_{dip}|++\rangle = -2(g_J J_z \beta)^2/r_{Ab}^3$ = +3.2 cm$^{-1}$ (For a $-2J$ Hamiltonian the value has to be divided by $-1/2$), where $g_J = 3/2$, $J_z = 6$, $r_{AB} = 3.5230$ Å.

# F. Supplementary Figures

1. ## MS

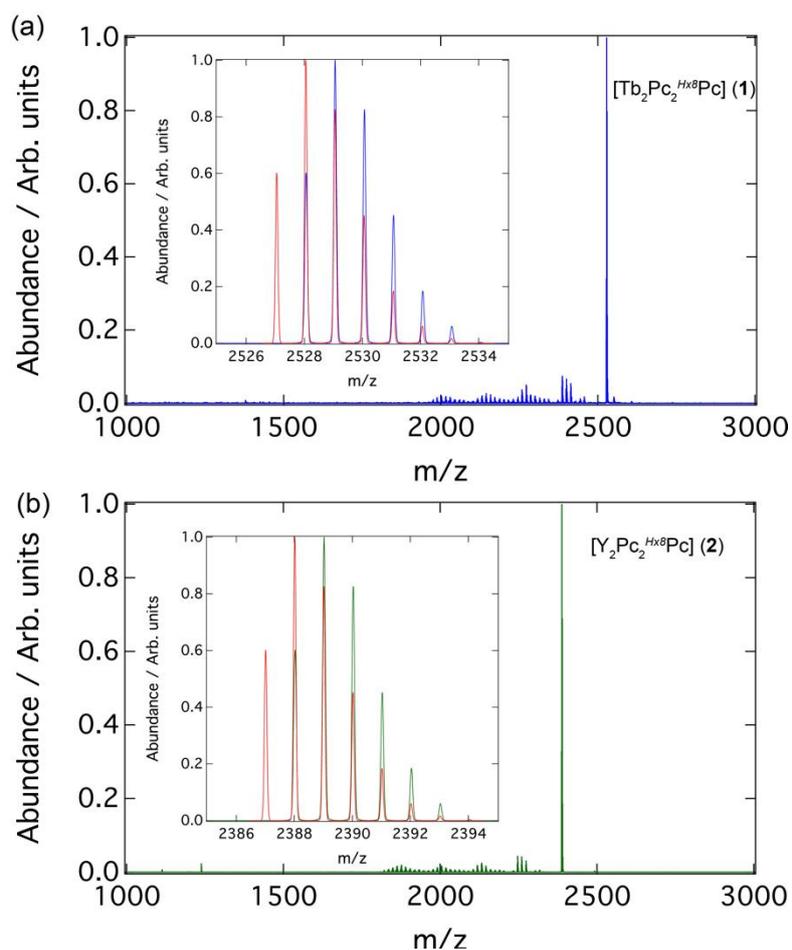

**Figure S1**. (a) MALDI mass spectra for compound **1** (a) and **2** (b). All data was collected employing positive ionization mode. Red traces correspond to the simulated MS pattern.

2. ## UV

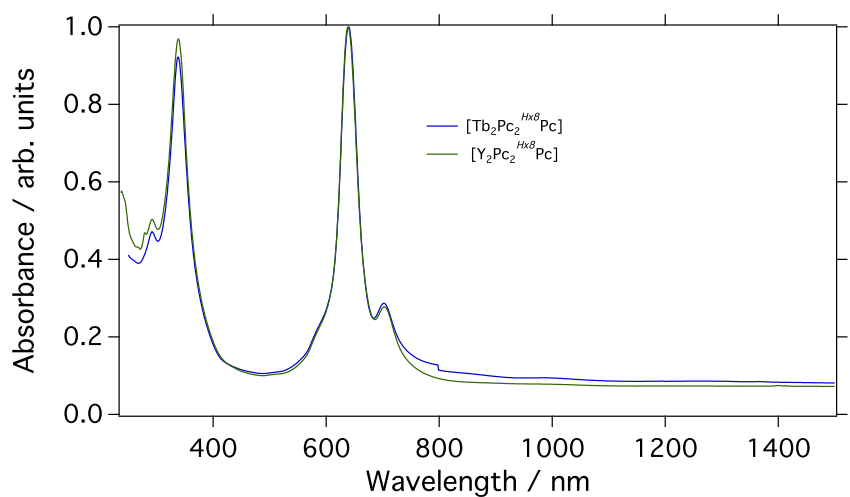

**Figure S2.** UV-vis spectra for compounds **1** and **2** in DCM.

## 3. Magnetic measurements

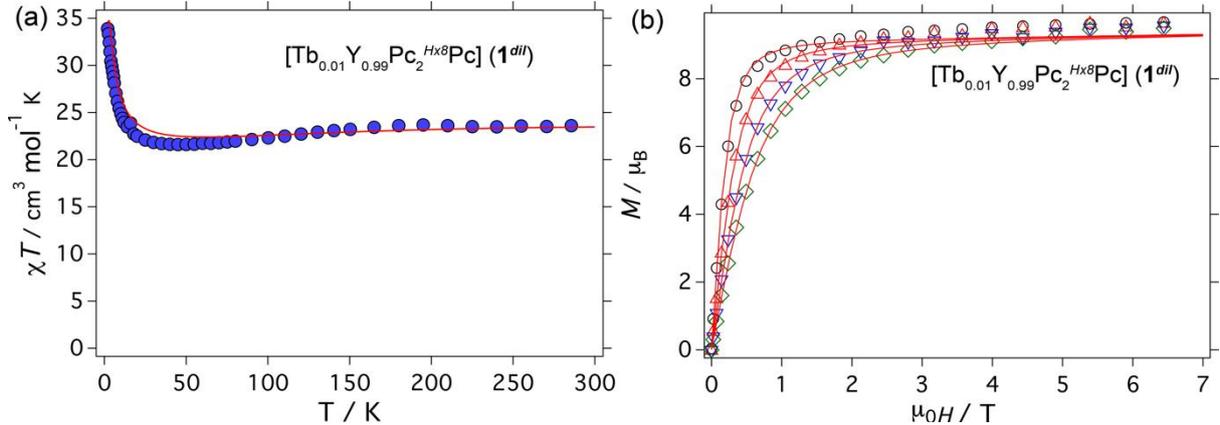

**Figure S3**. (a) $\chi_M T(T)$ and (b) $M(H)$ data for compound **1**$^{dil}$ and simulation (solid lines) using Hamiltonian (4) and parameters described in the text.

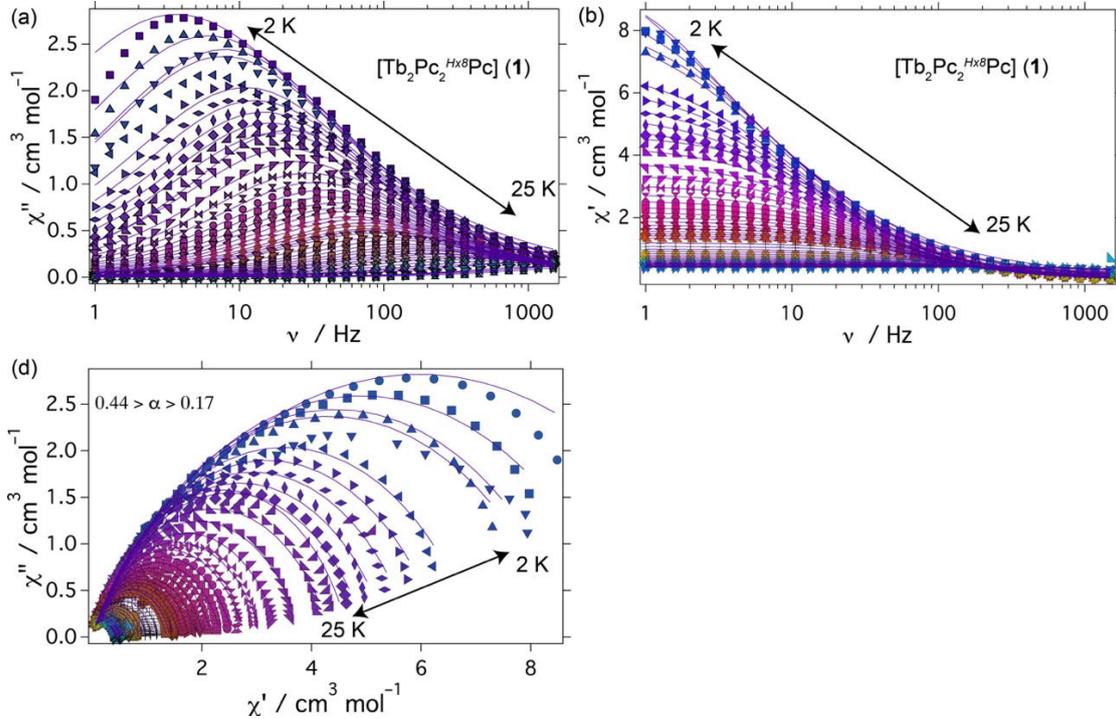

**Figure S4**. Dynamic magnetic data for compound **1**$^{dil}$ at zero DC field with an oscillating field of 3.5 Oe. Solid lines are fits to a single relaxation process. (a) $\chi_M'(\nu)$, (b) $\chi_M''(\nu)$ and (c) Cole-Cole traces.

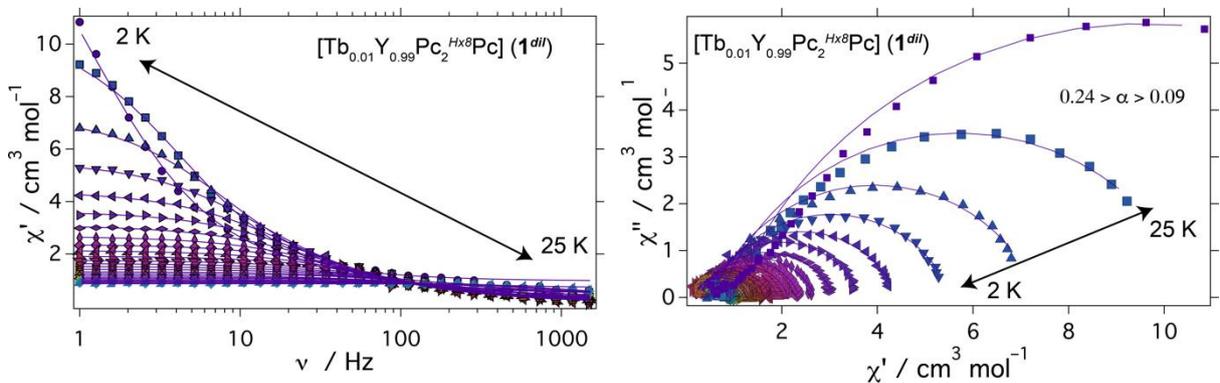

**Figure S5**. Dynamic magnetic data for compound **1**$^{dil}$ at zero DC field with an oscillating field of 3.5 Oe. Solid lines are fits to a single relaxation process. (a) $\chi_M'(\nu)$ and (b) Cole-Cole traces.

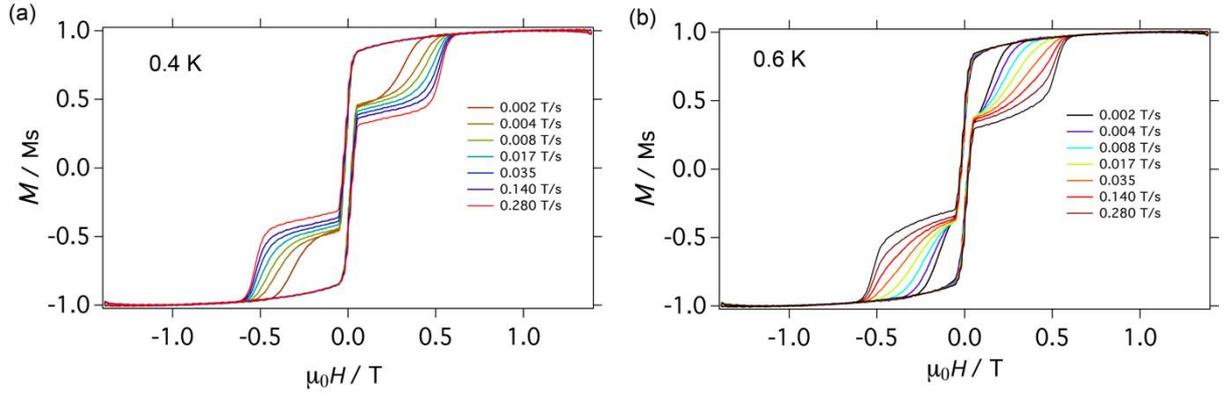

**Figure S6.** µ-SQUID data of **1**<sup>*dil*</sup> obtained at different field sweep rates at (a) 0.4 K and (b) 0.6 K. First derivative of the magnetization of µ-SQUID loops shown at a fixed sweeping rate (0.140 T/s) and different temperatures is given in Figure S8.

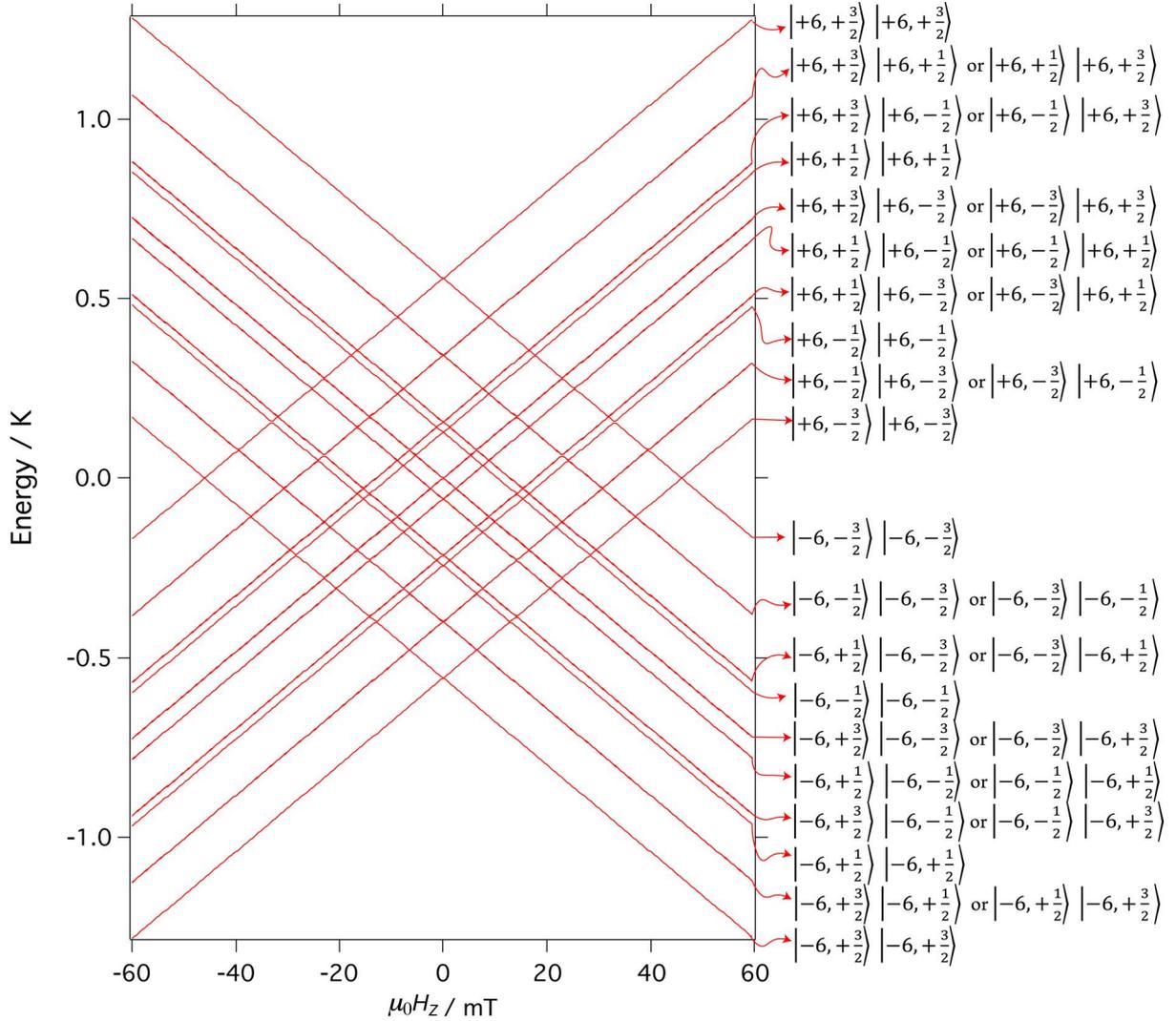

**Figure S7.** Zeeman diagram employing Hamiltonian (3) and parameters described in the MS.

S9

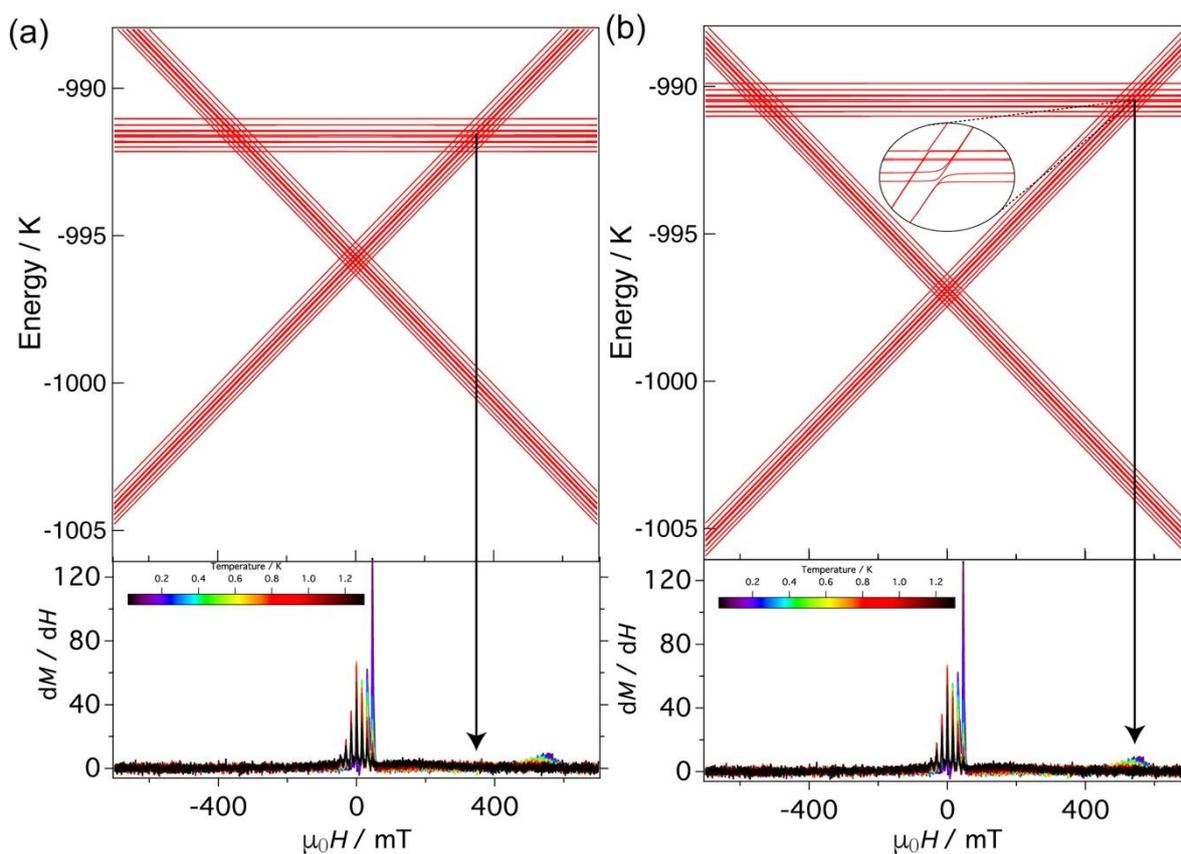

**Figure S8.** (a) Simulated Zeeman diagram using Hamiltonian (3) with $\mathcal{H}_{dip}$= +0.0223 cm$^{-1}$ and (b) with $\mathcal{H}_{dip} + J_{Ex}$, where $J_{Ex}$ = +0.0097 cm$^{-1}$. In (a) the arrow shows the absence of the transition at ±430 mT employing solely a dipole-dipole argument, whilst (b) shows the agreement with the transition at ±550 mT. The avoided crossing (zoomed region) arises from single spin flipping between the states $|\pm6, \pm1/2\rangle|\mp6, \mp1/2\rangle \leftrightarrow |\pm6, \pm1/2\rangle|\pm6, \mp1/2\rangle$ or $|\mp6, \mp1/2\rangle|\pm6, \pm1/2\rangle \leftrightarrow |\mp6, \mp1/2\rangle|\mp6, \pm1/2\rangle$.